\documentclass[12pt]{article}

\usepackage{amssymb}
\usepackage{amsmath}
\usepackage{latexsym}

\catcode`@=11
\renewcommand{\section}{\@startsection{section}{1}{0pt}{\medskipamount}
{\medskipamount}{\large\bf}}
\numberwithin{equation}{section}
\catcode`@=12

\def\beq{\begin{eqnarray}}    
\def\eeq{\end{eqnarray}}      

\def\pa{\partial}                       

\def\={\ =\ }

\def\vep{\varepsilon}

\begin{document}

\begin{center}

{\Large\bf Conversion of second-class constraints and
resolving the zero curvature conditions
in the geometric quantization theory}

\vspace{18mm}

{\large Igor A. Batalin$^{(a,b)}\footnote{E-mail:
batalin@lpi.ru}$\;, Peter M. Lavrov$^{(b, c)}\footnote{E-mail:
lavrov@tspu.edu.ru}$\; }

\vspace{8mm}

\noindent ${{}^{(a)}}$
{\em P.N. Lebedev Physical Institute,\\
Leninsky Prospect \ 53, 119 991 Moscow, Russia}

\noindent  ${{}^{(b)}}
${\em
Tomsk State Pedagogical University,\\
Kievskaya St.\ 60, 634061 Tomsk, Russia}

\noindent  ${{}^{(c)}}
${\em
National Research Tomsk State  University,\\
Lenin Av.\ 36, 634050 Tomsk, Russia}

\vspace{20mm}

\begin{abstract}
\noindent
 In the approach to geometric quantization based on the conversion of
second-class constraints, we resolve the corresponding nonlinear
zero-curvature conditions for the extended symplectic potential.
From the zero-curvature conditions, we deduce new linear equations
for the extended symplectic potential. We show that solutions of the
new linear equations also satisfy the zero-curvature condition. We
present a functional solution of these new linear equations and
obtain the corresponding path integral representation. We
investigate the general case of a phase superspace where boson and
fermion coordinates are present on an equal basis.
\end{abstract}

\end{center}

\vfill

\noindent {\sl Keywords:} Symplectic potential, second-class constraints,
conversion method
\\

\noindent PACS numbers: 11.10.Ef, 11.15.Bt
\newpage

\section{Introduction and summary}

We are happy for the opportunity to contribute an article in honor
of the 75th birthday of our friend and colleague Professor Igor
Viktorovich Tyutin. We have worked together with Igor Tyutin for
many years, and we have valued his extraordinary scientific
potential in full measure, as well as his brilliant personal
qualities. We cordially wish Igor Tyutin many new scientific
achievements together with further successes in his personal life.

Berezin's fundamental concept of quantization~\cite{Ber} has a
nontrivial projection on the geometric
quantization~\cite{SW,C,FH,W}, the Batalin--Fradkin--Vilkovisky
(BFV) formalism~\cite{FV,BVhf}, and the deformation
quantization~\cite{Fed,AKSZ,CF}. It is well known that the
conversion of second-class constraints serves as one of the most
powerful modern approaches to geometric quantization. The standard
scenario of the conversion method~\cite{BF2,BF1,BFF1,BF3,BFF3,FL1}
(also see~\cite{FL} and
further developments in~\cite{GL,BGL1,BGL}) is formulated as follows. The
starting point is some phase manifold with a complicated nonlinear
Poisson bracket. New momenta are then introduced as canonically
conjugate to the original phase variables, which are now regarded as
mutually commuting. Second-class constraints are simultaneously
introduced equating the new momenta to the components of the
symplectic potential. Additional degrees of freedom (conversion
variables)~\cite{BF3,BF4} are then introduced to convert the
second-class constraints into first-class constraints. The Poisson
brackets for the conversion variables are defined to be constant.
The first-class constraints obtained after the conversion equate the
momenta to the components of the extended symplectic potential,
which now also depend on the conversion variables. It is usually
assumed that these first-class constraints mutually commute (Abelian
conversion). Their commutation relations then have the form of a
zero-curvature condition. These zero-curvature conditions are in
fact nonlinear equations that are technically difficult to solve for
the extended symplectic potential.

Here, we derive new linear equations for the extended symplectic potential
by multiplying the zero-curvature conditions times the original phase
variables. We then show that solutions of these linear equations also
satisfy nonlinear zero-curvature conditions. The situation here is
very similar to the case of the Maurer--Cartan equation in standard group
theory. Finally, we present a functional solution of the new linear
equations and then derive the corresponding path integral representation.

\section{Classical mechanics in general setting}

Let   $Z^{A}$,  $\varepsilon( Z^{A} )  =  \varepsilon_{A}$,  be coordinates of
original phase space within the Hamiltonian formalism.
Let  $V_{A}( Z )$,  $\varepsilon( V_{A} ) = \varepsilon_{A}$,  be a  symplectic
potential,  and  $S$  be an action, with an original Hamiltonian $H=H( Z )$,
\beq
\label{n1}
S   =  \int  dt  \mathcal{ L },   \quad
\mathcal{ L }  =  V_{A} \pa_{t}Z^{A}  -  H.  
\eeq
Making an arbitrary variation $\delta Z^A$, we obtain the equations of
motion
\beq
\label{n2}
\omega_{AB} \pa_{t} Z^{B}  -  \pa_{A} H   =  0,   
\eeq
where $\omega_{AB}$  is  a  symplectic  metric,
\beq
\label{n3}
\omega_{AB}  =  \pa_{A} V_{B} + \pa_{B} V_{A} (-1)^{ ( \varepsilon_{A} + 1 ) (
\varepsilon_{B} + 1 ) }.     
\eeq
It follows from (\ref{n3}) that
\beq
\label{n0.4}
\pa_{C} \omega_{AB} (-1)^{ ( \varepsilon_{C}  + 1 ) \varepsilon_{B} }  +
{\rm cyclic \;perm.}\; ( A, B, C )  =  0.    
\eeq
Hereafter, we assume that metric~(\ref{n3}) is invertible and let
\beq
\label{n4}
\omega^{AB} =  -  \omega^{BA} (-1)^{ \varepsilon_{A} \varepsilon_{B} },   
\eeq
denote its inverse. From~(\ref{n0.4}), we hence obtain
\beq
\label{n0.6}
\omega^{AD} \pa_{D} \omega^{BC} (-1)^{ \varepsilon_{A} \varepsilon_{C} }  +
{\rm cyclic \;perm.}\; ( A, B, C ) =  0.  
\eeq
Multiplying~(\ref{n3}) by $Z^A$ from the left, we obtain
\beq
\label{n0.7}
 ( Z^{A} \pa_{A} + 1 ) V_{B}  - \mathcal{F}_{0}\overleftarrow{\pa}_{B}
 =  Z^{A}\omega_{AB},   
\eeq
for $V_B$ with an arbitrary function $\mathcal{F}_0=Z^AV_A(-1)^{\vep_A}$. In turn,
it follows from~(\ref{n0.7}) that the Lagrangian in~(\ref{n1}) can be rewritten in the
form
\beq
\label{n0.9}
\mathcal{L}  =  Z^{A} \bar{\omega}_{AB} \pa_{t} Z^{B} - H + \pa_{t} \chi,  
\eeq
where $\chi$  is an arbitrary function and the barred metric $\bar{ \omega }$
is defined by the equation
\beq
\label{n0.10}
( Z^{C} \pa_{C}  +  2 ) \bar{\omega}_{AB}  =  \omega_{AB}.    
\eeq
It then follows from~(\ref{n2}) that
\beq
\label{n5}
\pa_{t} Z^{A}  -  \omega^{AB} \pa_{B} H  =  0. 
\eeq
Now,  let  $P_{A}$,  $\varepsilon( P_{A} )  =  \varepsilon_{A}$,  be the
momenta  conjugate to  $Z^{A}$. Action~(\ref{n1}) is then rewritten in the form
\beq
\label{n6}
\mathcal{ L } =  P_{A} \pa_{t} Z^{A}  -  H  -  \Theta_{A} \lambda^{A}, 
\eeq
where $\lambda^{A}$,  $\varepsilon( \lambda^{A} )  =  \varepsilon_{A}$,  are
Lagrange  multipliers and
\beq
\label{n7}
\Theta_{A}  =   P_{A}  -  V_{A} ,    
\eeq
are second - class constraints,
\beq
\label{n8}
\{ \Theta_{A},  \Theta_{B} \}   =  \omega_{AB} (-1)^{ \varepsilon_{A} },
\quad \{  Z^{A},  P_{B} \}  =  \delta^{A}_{B}. 
\eeq
For any functions $F(Z)$ and $G(Z)$, we define the Dirac bracket related to
second-class constraints~(\ref{n7}),
\beq
\label{nv0.9}
\{ F, G \}_{D} = \{ F, G \} - \{ F, \Theta_{A} \}\;
\omega^{AB} (-1)^{\varepsilon_{B}} \{\Theta_{B}, G \},     
\eeq
where we use~(\ref{n8}). Because $\{F(Z),G(Z)\}=0$, it follows immediately
from~(\ref{nv0.9}) that
\beq
\label{nv0.10}
\{ F, G \}_{D} = F \overleftarrow{\pa}_{A}\; \omega^{AB}\; \overrightarrow{\pa}_{B} G.    
\eeq
Because of  (\ref{n0.6}), we have
\beq
\label{n0.11}
\{ F,  \{G, H \}_{D} \}_{D}  (-1)^{ \varepsilon_{F} \varepsilon_{H} }  +
{\rm cyclic \;perm.}\; ( F, G, H )  =  0.
\eeq
Dirac bracket~(\ref{nv0.10}) thus reproduces the ``curvilinear" Poisson bracket
related to metric~(\ref{n4}). In terms of Dirac bracket~(\ref{nv0.10}), equations of
motion~(\ref{n5}) are rewritten as
\beq
\pa_{t} Z^{A} = \{ Z^{A}, H \}_{D}.   
\eeq

So far, the vorticity  in the right-hand side
of (\ref{n3})  is  Abelian.  Now  let  $\Phi^{A}$, $\varepsilon( \Phi^{A} )  =
\varepsilon_{A}$,  be  the conversion variables whose
nonzero Poisson brackets are defined by
\beq
\label{n9}
\{ \Phi^{A},  \Phi^{B} \}  =  \eta^{AB}  =  {\rm const},     
\eeq
We assume that metric~(\ref{n9}) is invertible, and the inverse is denoted by
$\eta_{AB}$. In the spirit of the general ideology of the conversion method,
we define the new constraints
\beq
\label{n10}
T_{A}  =  P_{A}  -  {\cal V }_{A}( Z, \Phi  ),         
\eeq
and require that they be Abelian first-class constraints,
\beq
\label{n11}
\{ T_{A},   T_{B}\}  =  ( \pa_{A} {\cal V}_{B}   +
\pa_{B}{\cal V}_{A} (-1)^{ ( \varepsilon_{A} + 1 ) ( \varepsilon_{B} +1 ) }  )
(-1)^{\varepsilon_{A} }  +
\{{\cal V}_{A},  {\cal V}_{B} \}  =  0.       
\eeq
These equations should be solved for ${\cal V}_{A}(Z,\Phi)$ under the boundary
condition
\beq
\label{n12}
{\cal V}_{A}( Z, \Phi  )\Big|_{ \Phi = 0 }  =  V_{A}(  Z ).    
\eeq
In the linear order in $\Phi^A$,
\beq
\label{nv12}
{\cal V}_{A}   =   V_{A}   -   V_{AB} \Phi^{B}   +   \mathcal{O}( ( \Phi
)^{2} ),   
\eeq
and the  involution  relation  (\ref{n11})  yields
\beq
\label{nv13}
V_{AC}\; \eta^{CD}\; V_{BD} (-1)^{  ( \varepsilon_{B} + 1 ) \varepsilon_{D} }
= -  \omega_{AB}  (-1)^{ \varepsilon_{A} } .    
\eeq
Hence,  $V_{AB}$  serves  as  the  "matrix-valued square root"  of  the  metric
(\ref{n8}).
Higher-order coefficients of an expansion of ${\cal V}_{A}$ in powers of $\Phi$
serve as the higher-order structure functions of the bundle generated by
$V_{AB}$. ``Square root" equation~(\ref{nv13}) naturally defines the covariant
constancy of metric~(\ref{n3}),
\beq
\label{nv015}
\nabla_C( \Delta )\; \omega_{AB} = 0 = \pa_{C} \omega_{AB} -
\Delta_{CA}^{\;\;\;\;\;D} \;\omega_{DB} (-1)^{\varepsilon_{A} + \varepsilon_{D} } -
 \Delta_{CB}^{\;\;\;\;\;D}\; \omega_{AD} (-1)^{ \varepsilon_{A} ( \varepsilon_{B} +
\varepsilon_{D} ) },  
\eeq
where the remote parallelism connection is defined in terms of $V_{AB}$ and
its inverse $V^{AB}$ by
\beq
\label{nv0.16}
\Delta_{CA}^{\;\;\;\;\;D} = (\pa_{C} V_{AB} ) V^{BD}.    
\eeq
The corresponding curvature is equal to zero,
\beq
\nonumber
&&\pa_{E} \Delta_{CA}^{\;\;\;\;\;D}   -
\pa_{C} \Delta_{EA}^{\;\;\;\;\;D}  (-1)^{ \varepsilon_{E}\varepsilon_{C} }   +\\
&&+\Delta _{CA}^{\;\;\;\;\;F} \Delta_{EF}^{\;\;\;\;\;D}  (-1)^{ \varepsilon_{E} (
\varepsilon_{C} + \varepsilon_{A} + \varepsilon_{F} ) }  -
 \Delta_{EA}^{\;\;\;\;\;F} \Delta_{CF}^{\;\;\;\;\;D}  (-1)^{ \varepsilon_{C} (
\varepsilon_{A} + \varepsilon_{F} ) }   =   0,      
\eeq
and connection~(\ref{nv0.16}) is therefore integrable.

We seek a solution of Eq.~(\ref{n11}) in the form of an expansion in $\Phi$,
\beq
\label{nv015}
T_{A}  =  \sum_{n = 0}^{\infty} T^{ (n) }_{A}, \quad    T^{ (n) }_{A} \;\sim\;
  \Phi^{ n },  
\eeq
\beq
\label{nv016}
T^{ (0) }_{A}  =  P_{A}  -  V_{A}, \quad
T^{ (n) }_{A}  =  -  \mathcal{V}^{ (n)}_{A}, \quad  n \geq   1.   
\eeq
The recurrence relation for that solution follows from a theorem in~\cite{BTet}:
\beq
\label{nv017}
T^{ ( n+1) }_{A}  =  - (  n + 2 )^{-1}  \Phi^{D}  \eta_{DC}  V^{CB}  W^{ (n)
}_ {BA}  +
\{  T^{ (1) }_{A},  \mathcal{ K }^{ (n+2) } \}_{(\Phi)},   \quad n \geq  1, 
\eeq
where we introduce the notation
\beq
\label{nv018}
\mathcal{ K }^{ (n +2) } \;\sim\;  \Phi^{ (n+2) },    
\eeq
\beq
\label{2.28}
W_{AB}^{ (1) }  =  \{ T_{A}^{ (0) } ,  T_{B}^{ (1) } \}_{ (P, Z) }  -  \{
T_{B}^{ (0) } , T_{A}^{ (1) } \}_{ (P, Z) }  (-1)^{ \varepsilon_{A}
\varepsilon_{B} },     
\eeq
\beq
\label{nv019}
W^{ (n) }_{AB}  =  \sum_{m =  0}^{n}  \{  T^{ (n-m) }_{A},   T^{ (m) }_{B}
\}_{ ( P, Z ) }   +
\sum_{m = 0}^{ (n-2) }  \{  T^{ (n-m) }_{A} ,   T^{ (m+2) }_{B}  \}_{ (
\Phi ) },    \quad   n \geq  2.   
\eeq
Now,  $\eta_{AB}$ and $V^{AB}$ are respectively inverse to $\eta^{AB}$ and
$V_{AB}$. Moreover, we let $\{ \; ,  \}_{ (P,Z) }$ and $\{ \; ,  \}_{ ( \Phi ) }$
denote the corresponding Poisson brackets in the $P$, $Z$, and $\Phi$
sectors. A particular example of recurrence relation~(\ref{nv017}) at $n=1$ has the
form
\beq
\label{nv020}
T^{ (2) }_{A}  =  - \frac{1}{3}  \Phi^{D}  \eta_{DC}  V^{CB}  \big(  \{  P_{B}, V_{AE}
\}  -\{  P_{A}, V_{BE}\}   (-1)^{ \varepsilon_{A} \varepsilon_{B} } \big)  \Phi^{E}
-\{  \mathcal{ V }^{ (1) }_{A} ,  \mathcal{ K }^{ (3) }  \}_{ (\Phi) }.
\eeq

The vorticity in the right-hand side of~(\ref{n11}) definitely has the structure
characteristic for the Yang--Mills theory, and curvature tensor
equation~(\ref{n11}) is hence a typical zero-curvature condition. Therefore, the
solution for ${\cal V}_{A}$ has the usual structure of the Cartan form, which in
turn generates non-Abelian ordered exponentials.

We now consider the extended Hamiltonian $\mathcal{H}$ defined by
\beq
\label{2.30}
\{\mathcal{H}, T_{A}\}=0,  \quad  \mathcal{H} \big|_{\Phi = 0}  =  H , 
\eeq
where $T_A$ are Abelian constraints~(\ref{n10}) after conversion. Just as
in~(\ref{nv015}), we seek a solution of~(\ref{2.30}) in the form of an expansion in
$\Phi$,
\beq
\label{2.31}
\mathcal{H}  =  \sum_{n = 0}^{\infty}\mathcal{H}^{(n)}  ,\quad
 \mathcal{H}^{(n)}  \sim   \Phi^{n},   \quad   \mathcal{H}^{(0)} = H. 
\eeq
The recurrence relation for that solution also follows from the theorem
in~\cite{BTet}:
\beq
\label{2.32}
\mathcal{H}^{(n + 1)}  =  -  (n + 1)^{-1}\Phi^{A}\eta_{AB} V^{BC}W_{C}^{(n)},
\quad    n  \geq  0,   
\eeq
where we introduce the notation
\beq
\label{2.33}
W_{A}^{(0)}  =  \{T_{A}^{(0)} , \mathcal{H}^{(0)}\}, 
\eeq
\beq
\label{2.34}
W_{A}^{(1)}  =  \{T_{A}^{(1)} , \mathcal{H}^{(0)}\}  +
\{T_{A}^{(0)} ,  \mathcal{H}^{(1)} \}  +
\{T_{A}^{(2)} , \mathcal{H}^{(1)}  \}_{(\Phi) },  
\eeq \beq \label{2.35} W_{A}^{(n)}  =  \sum_{m = 0}^{n} \{ T_{A}^{(n
- m)} ,  \mathcal{H}^{(m)}\}_{(P, Z)}   + \sum_{m = 0}^{n - 2}
\{T_{A}^{(n - m)},  \mathcal{H}^{ (m + 2) }\}_{ (\Phi)}  +
 \{T_{A}^{ (n + 1) } , \mathcal{H}^{(1)}\}_{ (\Phi) },   \quad  n  \geq 2. 
\eeq
In particular, for $n=0,1$, we thus obtain
\beq
\label{2.36}
\mathcal{H}^{(1)}  =  -  \Phi^{A} \eta_{AB}V^{BC} \{ P_{C} ,  H \},  
\eeq
\beq
\nonumber
&&\mathcal{H}^{ (2) }  =  -\frac{1}{2} \Phi^{A} \eta_{AB} V^{BC}  \Big[  -
\{ P_{C} ,\Phi^{D}  \eta_{DE}  V^{EF}  \{ P_{F} ,  H \} \}  +\\
\nonumber
&&
+\frac{1}{3} \Big( \Phi^{D} \eta_{DE} V^{EF} \big( \{ P_{F} , V_{CG} \}  -
\{ P_{C} ,V_{FG}\} (-1)^{ \varepsilon_{C} \varepsilon_{F} } \big)  -\\
&&-(-1)^{ \varepsilon_{C} \varepsilon_{E} } V^{EF} \big( \{ P_{F}, V_{CD} \}  -
\{ P_{C} , V_{FD} \} (-1)^{ \varepsilon_{C} \varepsilon_{F} } \big) \Phi^{D}
\eta _{EG} \Big) V^{GM} \{ P_{M},  H \}  \Big] .   
\label{2.37}
\eeq
We note that the extension of the original phase space
$( Z  ) \rightarrow  ( Z, P)\rightarrow ( Z, P, \Phi )$
can be formulated directly at the level of action~(\ref{n1}),
\beq
\nonumber
&&S \;\rightarrow\;  \int dt \left[  P_{A} \pa_{t} Z^{A}  -  H( Z )  -
 \Theta_{A}( Z, P )
\lambda^{A} \right]\rightarrow  \\
\nonumber
&&\int dt  \left[ P_{A} \pa_{t} Z^{A}  +
\frac{1}{2} \Phi^{A} \eta_{AB} \pa_{t} \Phi^{B}  -  \mathcal{ H }( Z, \Phi )  -
T_{A}( Z, P, \Phi )  \lambda^{A} \right] \;\rightarrow\; \\
&& \mathcal{ S }   =
\int dt  \left[ \mathcal{V}_{A} (Z, \Phi ) \pa_{t} Z^{A}  +\frac{1}{2}
\Phi^{A}\eta_{AB} \pa_{t} \Phi^{B}  -  \mathcal{ H }( Z, \Phi  ) \right],
\eeq
where $\lambda^A$ are Lagrange multipliers of the corresponding constraints.
The Abelian first-class constraints $T_A$ given by~(\ref{n10}) generate gauge
transformations leaving the action $\mathcal{ S }$ invariant,
\beq
\delta Z^{A}  =  \Xi^{A}( t ),   \quad  \delta \Phi^{A}   =   - \eta^{AB} ( \pa_{B}
\mathcal{V}_{C} ) \Xi^{C}( t ) ,   \quad  \pa_{B} = \frac{\pa}{ \pa \Phi^{B}}. 
\eeq

\section{Quantum description}

Our consideration extends to quantum mechanics by simply replacing the
Poisson brackets with (super)commutators. The last rule applies directly
to~(\ref{n8}), (\ref{n9}), (\ref{n11}), and so on as
\beq
\label{n13}
\{\; , \}\; \rightarrow\; ( i \hbar)^{-1} [\; , ].    
\eeq
Hence, the operator-valued analogue of~(\ref{n11}) is represented as the relation
\beq
\label{nv14}
( i \hbar )^{-1} [ T_{A}, T_{B} ]  =  (  \pa_{A} {\cal V}_{B} +
 \pa_{B} {\cal V}_{B} (-1)^{ ( \varepsilon_{A} + 1 ) (
\varepsilon_{B} + 1 ) }  ) (-1)^{\varepsilon_{A}}  +
( i \hbar )^{-1} [{\cal V}_{A}, {\cal V}_{B} ]  =  0,     
\eeq
\beq
\label{nv15}
{\cal V}_{A} \Big|_{ \Phi = 0 }   =   V_{A}.    
\eeq
We derive a linear differential equation that describes the structure of a
solution for $\mathcal{V}_{A}$. Multiplying Eq.~(\ref{nv14})
by $Z^{A} (-1)^{ \varepsilon_{A} }$ from the
left, we obtain (cf.~(\ref{n0.7}))
\beq
\label{0.30}
( Z^{A} \pa_{A}  +  1 ) \mathcal{V}_{B}   -  \pa_{B} {\cal F} (-1)^{ \varepsilon_{B}
}  +  ( i  \hbar )^{-1} [ {\cal F} , \mathcal{V}_{B} ]  =  0,   
\eeq
where we introduce the notation
${\cal F}  =  Z^{A} \mathcal{V}_{A} (-1)^{ \varepsilon_{A} }$. Rescaling the
variables $Z^A$ with a bosonic parameter~$t$,
\beq
\label{0.32}
Z^{A} \;\rightarrow\; t  Z^{A},     
\eeq
we rewrite (\ref{0.30}) in the form
\beq
\label{0.33}
\pa_{t} ( t  \mathcal{ V}_{B} )-\pa_{B}  F  (-1)^{ \varepsilon_{B} }   +
( i  \hbar  )^{-1}  [ F ,t \mathcal{V}_{B} ]=0,\quad
F  =  F( t,  Z, \Phi )  =  Z^{A} \mathcal{V}_{A}( t Z, \Phi ) (-1)^{
\varepsilon_{A} }.       
\eeq
It in turn follows from the last equation that
\beq
\label{*}
 t  \mathcal{V}_{B}  =   \int _{0}^{t}  dt' \; T  \exp\left\{\frac{i}{\hbar }
\int _{t'}^{t}  F  d t'' \right\} \left(F \overleftarrow{\pa}_{B}\right) \;  T^*
\exp\left\{ -\frac{i}{ \hbar} \int _{t'}^{t}  F  dt''  \right\}  
\eeq
where $T$ and $T^*$ respectively denote chronological and antichronological
ordering. Introducing external sources $J_A(t)$ to the Weyl-ordered
operators $\Phi^A$ ((\ref{nv16})), we represent the functional solution for
the left chronologically ordered exponential in~(\ref{*}) as
\beq
\nonumber
&&T  \exp\left\{\frac{i}{  \hbar}  \int _{t'}^{t}F( t'', Z,\Phi (t'') )d t'' \right\}=\\
\nonumber && \left[ \; \exp\left\{\frac{i}{\hbar }  \int_{t'}^{t}
F( t'', Z ,  (  \hbar / i  ) ( \delta
/ \delta J( t'' ) )  d t''  \right\}\right.\times\\
\nonumber && \times\exp\left\{-\frac{i}{ 4 \hbar }  \int _{t'}^{t}
d t''  \int_{t'}^{t}  d t''' J_{A}( t'' )  \eta^{AB}  {\rm sign}(
t'' - t''' ) J_{B}( t''' )(-1)^{\varepsilon_B} \right\}\times
\\
\label{0.38} && \times \left.\exp\left\{\frac{i}{\hbar }
\int_{t'}^{t}  J_{A}(t'')  d t''\Phi^{A}\right\}\right]
\Big|_{ J = 0 } .           
\eeq
Performing a functional Fourier transformation, we can derive a path
integral representation for the left-ordered exponential in~(\ref{*})
from~(\ref{0.38}). We thus obtain the path integral representation for
solution~(\ref{0.38}),
\beq
\label{0.39}
\int [ D \Gamma ] \exp\left\{\frac{i}{\hbar}  \int_{t'}^{t} d t''  \left[\frac{1}{2}
\Gamma^{A}\; \eta_{AB}\; \pa_{t''} \Gamma^{B}   +
F(  t'', Z,  \Gamma( t'' )  )  \right]+
\frac{i}{2 \hbar}  \Gamma^{A}( t )  \eta_{AB}  \Gamma^{B}( t' )  \right\},    
\eeq
where the virtual integration trajectories satisfy the boundary condition
\beq
\label{0.40}
\Gamma^{A}( t ) + \Gamma^{A}( t' )  =   2  \Phi^{A}.      
\eeq
Expression~(\ref{0.39}) should be regarded as a Weyl-ordered function of the
operators $\Phi^A$. To obtain the corresponding functional solution for the
right-ordered exponential in~(\ref{*}), we should change the sign of the
argument in the first and second exponential in the right-hand side
of~(\ref{0.38}).

We assert that every solution of Eq.~(\ref{0.33}) with an arbitrary $F$ also satisfies
Eq.~(\ref{nv14}). From~(\ref{0.33}), it is easy to derive the Cauchy problem
\beq
\label{0.34}
\pa_{t} X_{AB} = -( i \hbar )^{-1} [ F , X_{AB} ],   \quad
X_{AB}\Big|_{t = 0}  =0,   
\eeq
where
\beq
\label{0.35}
X_{AB}  =  (  \pa_{A} t \mathcal{V}_{B}  +
\pa_{B} t \mathcal{V}_{A} (-1)^{ ( \varepsilon_{A} + 1 ) (
\varepsilon_{B} + 1 ) }  ) (-1)^{ \varepsilon_{A} }  +
 ( i  \hbar )^{-1}  [ t  \mathcal{V}_{A} ,  t  \mathcal{V}_{B} ],  
\eeq
with the potential $\mathcal{V}_{A}$ rescaled by~(\ref{0.32}).
By standard arguments, it
follows from~(\ref{0.34}) that
\beq
\label{0.36}
X_{AB}  =  0, \quad   {\rm for \;any}\; t,     
\eeq
and at $t=1$, we hence obtain Eq.~(\ref{nv14}). The operator $F$ describes the
arbitrariness of the solution.

Here and hereafter, we assume that a definite type of normal
ordering is chosen for all operator-valued functions. For
definiteness, we choose the Weyl ordering. The corresponding star
multiplication of the Weyl symbols is then given by
 \beq
\label{nv16} \star=\exp\left\{ \frac{i \hbar}{2
}\;\overleftarrow{\pa}_{A}\; \eta^{AB}\;
\overrightarrow{\pa}_{B}  \right\}, \quad  \pa_{A} = \frac{\pa}{\pa\Phi^{A}},   
\eeq
\beq
\label{nv17}
( {\rm Operator}\; F ) ( {\rm Operator}\; G )\;\rightarrow \;
( {\rm Symbol}\; F ) \star ( {\rm Symbol}\; G ), 
\eeq
\beq
\label{nv18}
( {\rm Operator}\; F )=
\exp\left\{ ( {\rm Operator}\; \Phi^{A} ) \pa_{A}\right\}
( {\rm Symbol}\; F )\Big|_{ \Phi =  0 }.  
\eeq

In the quantum mechanical setting, the complete unitarizing Hamiltonian
after conversion is given by
\beq
\label{n14}
H_{\Psi} = \mathcal{H} + (i \hbar)^{-1} [ \Psi, \Omega ].  
\eeq
where  $\mathcal{H}$ is a solution to the Cauchy problem
\beq
\label{n15}
( i  \hbar )^{-1} [ \mathcal{H},  T_{A} ]  =\mathcal{H}\overleftarrow{\pa}_{A} -
(i \hbar)^{-1} [ \mathcal{H}, {\cal V}_{A} ] = 0, 
\eeq
\beq
\label{n16}
\mathcal{H} |_{\Phi = 0} = H. 
\eeq
The gauge fermion has the standard form
\beq
\label{n17}
\Psi = \bar{C}_{A} \chi^{A} + \bar{\mathcal{P}}_{A} \lambda^{A},  
\eeq
and the BRST-BFV generator is typical as to the case of Abelian constraints,
\beq
\label{n18}
\Omega = C^{A} ( P_{A} - {\cal V}_{A} ) + \pi_{A} \mathcal{P}^{A},\quad
[ \Omega, \Omega ]  =  0 ,  
\eeq
under the standard canonical commutation relations in relativistic phase space,
\beq
\label{n19}
[ \lambda^{A}, \pi_{B} ] = [ C^{A}, \bar{\mathcal{P}}_{B} ] =
[ \mathcal{P}^{A}, \bar{C}_{B} ] = i \hbar \delta^{A}_{B},  
\eeq
\beq
\label{nv033}
\varepsilon(\lambda^{A}) = \varepsilon(\pi_{A}) = \varepsilon_{A},\quad
\varepsilon(C^{A}) = \varepsilon(\bar{\mathcal{P}}_{A}) =
\varepsilon(\mathcal{P}^{A}) = \varepsilon(\bar{C}_{A}) = \varepsilon_{A} + 1. 
\eeq

We now consider involution equation~(\ref{n15}) analogously to what we previously
did with Eq.~(\ref{nv14}). Multiplying~(\ref{n15}) by $Z^A$ from the right and then
doing rescaling~(\ref{0.32}), we derive
\beq
\label{3.25}
 \pa_{t} \mathcal{H}  -  ( i \hbar )^{-1} [ \mathcal{H},  F ] = 0, 
\eeq
where $F$ is defined  in (\ref{0.33}).
We assert that relations~(\ref{0.33}) and~(\ref{3.25})
lead to $\mathcal{H}$ that satisfies~(\ref{n15}) with an arbitrary $F$. From Eqs.~(\ref{0.33})
and~(\ref{3.25}), it is easy to derive the Cauchy problem, similar to~(\ref{0.34}),
\beq
\label{3.26}
\pa_{t} X_{A} = - ( i  \hbar )^{-1} [ F,  X_{A} ], \quad X_{A} |_{t = 0 } =  0.   
\eeq
where
\beq
\label{3.27}
X_{A} =    \mathcal{H}\overleftarrow{\pa}_{A}   -( i  \hbar )^{-1}
[  \mathcal{H},  t  \mathcal{V}_{A} ] .      
\eeq
By  standard argument,  it follows from (\ref{3.27}) that
\beq
\label{3.28}
X_{A} = 0,  \quad   {\rm for \;any}\; t,        
\eeq
and at $t=1$, we hence obtain (\ref{n15}). Similar reasonings are applicable
to any observable similar to $\mathcal{H}$.

We note that the rescaling procedure applied to~(\ref{nv14}) and (\ref{n15}) can be
naturally generalized as follows. Instead of multiplying the corresponding
equation by $Z^A$ and then using rescaling~(\ref{0.32}), we consider the line
\beq
\label{n3.29}
\pa_{s} \bar{Z}^{A}  =  \chi^{A}( \bar{Z} ),   \quad   \bar{Z}^{A}( s = 0 )  =
Z^{A},  \quad  \pa_{s}  =  \frac{\pa}{ \pa s},     
\eeq
where  $\chi^{A}( \bar{Z} )$ are some regular functions.
For example, if we choose
$\chi^A=\bar{Z}^{A}$, then~(\ref{n3.29}) yields
\beq
\label{n3.30}
\bar{Z}^{A}  =  Z^{A}  \exp\{ s \},   
\eeq
which is just the rescaling (\ref{0.32}) with  $t  =  \exp\{ s \}$.
We can now use general
line~(\ref{n3.29}) to define the generalized ``rescaling" procedure for the
corresponding equations. Namely, we should take the equation at the point
$\bar{Z}$ and then multiply by $\pa_{s} \bar{Z}$. For example, applied to
Eq.~(\ref{nv14}), the generalized rescaling procedure at $\bar{Z}$ yields
\beq
\label{n3.31}
( \pa_{s} \mathcal{V}_{A} G^{A}_{B} )   -  F \overleftarrow{\pa}_{A} G^{A}_{B}  +
( i  \hbar )^{-1}  [ F,  \mathcal{V}_{A} G^{A}_{B} ]  =  0,  
\eeq
where $G^{A}_{B}$ and $F$ are defined as
\beq
\label{n3.32}
\pa_{s} G^{A}_{B}  =  \chi^{A} \overleftarrow{\pa}_{C} G^{C}_{B}  , \quad  G^{A}_{B}( s
= 0 )  =  \delta^{A}_{B}, \quad    F  =  \mathcal{V}_{A} \chi^{A}.     
\eeq
It can be shown that generalized linear equation (\ref{n3.31}) implies the
corresponding counterpart of Cauchy problem (\ref{0.34}).
Hence, if~(\ref{n3.31}) is
satisfied, then its solutions also satisfy~\ref{nv14}).

In conclusion, we mention that the representation of the solution for
${\cal V}_{A}$ in the Cartan form has the natural realization
\beq
\label{n20}
{\cal V}_{A} =(-i\hbar) ( \pa_{A} U ) U^{-1} (-1)^{\varepsilon_{A}}.     
\eeq
It is easy to verify that every solution of~(\ref{n20}) satisfies~(\ref{nv14}). If we
solve~(\ref{n20}) for $U$, then we obtain the $T$-ordered exponential
\beq
\label{n21}
U = T \exp \left\{\frac{i}{\hbar} \int {\cal V}_{A} d Z^{A} \right\}.   
\eeq
From~(\ref{n20}), we then derive the equation ($T^*$ denotes antiordering)
\beq
\label{nv19}
{\cal V}_{A}  =  T \exp\left\{\frac{i}{ \hbar} \int {\cal V} d Z\right\}
\overleftarrow{\pa}_{A}\;( - i \hbar )\;
T^* \exp\left\{ -\frac{i}{\hbar} \int {\cal V} d Z \right\}.   
\eeq
which is in fact  a contour-integral analogue to  zero curvature
equation (\ref{nv14}).

We rewrite Eq.~(\ref{nv14}) for symbols in the form
\beq
\label{n30} (  \pa_{A}  \mathcal{V}_{A}   +
   \pa_{B}  \mathcal{V}_{B} (-1)^{ ( \varepsilon_{A} + 1 ) ( \varepsilon_{B}
+ 1 ) }  ) (-1)^{ \varepsilon_{A} }   +
  (  i  \hbar  )^{-1}  (   \mathcal{V}_{A}  \star  \mathcal{V}_{B}   -
  \mathcal{V}_{B}  \star  \mathcal{V}_{A} (-1)^{ \varepsilon_{A}
\varepsilon_{B} }  )    =    0,   
\eeq
where the $\star$- multiplication is defined in (\ref{nv16}).
We can seek a solution of this equation in the form of a semiclassical expansion in powers
of $\hbar$. The lowest-order quantum correction to classical equation~(\ref{n11})
is
\beq
\label{n31}
\frac{1}{3} (i\hbar)^{2}\;\mathcal{V}_{A}\Delta^{3}\mathcal{V}_{B},  
\eeq
where we set
\beq
\label{n32}
2  \Delta  =  \frac{\overleftarrow{\pa}}{\pa \Phi^{A}}\; \eta^{AB}
\frac{\overrightarrow{\pa}}{\pa\Phi^{B}} .     
\eeq

Equation (\ref{n15}) rewritten  for symbols  has  the  form
\beq
\label{n33}
\mathcal{H} \overleftarrow{\pa}_{A}   - ( i  \hbar )^{-1} ( \mathcal{H}  \star
\mathcal{V}_{A}   -
\mathcal{V}_{A}  \star  \mathcal{H}  )  =  0.     
\eeq
At  $\hbar  =  0$,  we obtain
\beq
\label{n34}
\mathcal{H}\overleftarrow{\pa}_{A}  -  \{ \mathcal{H},  \mathcal{V}_A \}  =  0.  
\eeq
from  (\ref{n33}).
The lowest order quantum correction to classic equation (\ref{n34}) is
\beq
\label{n35}
- \frac{1}{3} (i\hbar)^{2}\;\mathcal{H}\;\Delta^{3}\mathcal{V}_{A}. 
\eeq

Formulas  (\ref{n31}) and (\ref{n35})   are  particular  cases of
the  specific form  of  the   Weyl   symbol  commutators,
\beq
\label{n36} (  i  \hbar  )^{-1}  [  F ,  G  ]_{\star}   =   ( i
\hbar  )^{-1}   (  F \star  G   -   G  \star  F  (-1)^{
\varepsilon_{F} \varepsilon _{G} }  )   =
 2  (  \hbar  )^{-1}   F  \sin( \hbar \Delta ) \; G,          
\eeq
which  follows  from  the  antisymmetry  property  of  the  operator  $\Delta$
given by (\ref{n32}).

As an explicit example of how symbol equations like~(\ref{n30}) and (\ref{n33}) work
to ensure the generating mechanism for the structure relations of the
converted constraint algebra, we present a more detailed procedure for the
expansion in $\hbar$ and $\Phi$ in the case of Eq.~(\ref{n30}). We start from the
expansion for $\mathcal{V}_{A}$ in powers of $\hbar$ and $\Phi$:
\beq
\label{1}
\mathcal{V}_{A}  =  \sum_{q=0}^{\infty} ( i \hbar )^{q}  \sum_{p =0}^{\infty}
V^{ ( q ) }_{A  B_{p} ... B_{1} } \Phi^{ B_{1} } ... \Phi^{ B_{p} },  
\eeq
The correponding form of the structure relations for $V_{A B_{p} ... B_{1} }$ :
\beq
\nonumber
&&\left(\pa_{A} V^{ ( q ) }_{ B B_{p} ... B_{1} }  +
\pa_{B} V^{ ( q ) }_{ A B_{p} ... B_{1} }  (-1)^{ ( \varepsilon_{A} + 1 )
( \varepsilon_{B} + 1 ) }  \right)  (-1)^{ \varepsilon_{A} }  +\\
\nonumber
&& +\sum_{ k, l, m, n, s \geq  0 }  \delta^{q}_{ k + l + s - 1}
\delta^{p}_{m + n} \frac{ (m + s)! (n+ s)! }{m ! s! n!}  ( 2 )^{ - s}\times
\\
\nonumber
&&\qquad\times Sym_{ (  B_{p} ... B_{1} ) } \left[  \left(  V^{ ( k ) }_{ A B_{p} ... B_{n+1} C_{s}
... C_{1} }  V^{ ( l ) C_{1} ... C_{s}}_{B\;\;\;\;\;\;\;\;\;\;\;\;\; B_{n} ... B_{1}}
(-1)^{\varepsilon_{B} \sum_{j = n + 1 }^{p} \varepsilon_{B_{ j }}}\right)\right. -\\
&&\qquad\qquad\qquad\qquad\quad -
( A \leftrightarrow B )(-1)^{ \varepsilon_{A} \varepsilon_{B}}\Big]  =  0 ,  
\eeq
where we introduce the notation
\beq
\label{3}
V^{( l ) C_{1} ... C_{s}}_{B\;\;\;\;\;\;\;\;\;\;\;\; B_{n} ... B_{1}}  =  \eta^{ C_{1}
D_{1} } ... \eta^{ C_{s} D_{s} }  V^{ ( l ) }_{ B D_{s} ... D_{1} B_{n} ...B_{1} }
(-1)^{\sum_{ i = 1 }^{s}  \varepsilon_{ D_{ i } } \left(  \sum_{ j  =  i + 1 }^{s}
\varepsilon_{ C_{ j } } + \varepsilon_{B}+ 1\right)}.      
\eeq
The symmetrization operation is defined by
\beq
\label{4}
&&Sym_{( B_{p} ...B_{1})}  [ X_{ B_{p} ... B_{1} } ]  =   X _{ A_{p} ...A_{1} }
S^{ A_{1} ...  A_{p} }_{ B_{p} ...  B_{1} },   
\eeq
\beq
\label{5}
p ! \; S^{ A_{1} ... A_{p} }_{ B_{p} ... B_{1} }  =  \Phi^{ A_{1} } ... \Phi^{
A_{p} } \overleftarrow{\pa}_{ B_{p} } ... \overleftarrow{\pa}_{B_{1}}, \quad   
\pa_{A} = \frac{\pa}{\pa \Phi^{A}}.
\eeq
In the case of Eq.~(\ref{n33}), the procedure for expanding in $\hbar$ and $\Phi$
is completely analogous (up to some elementary modifications) to what was
done in the case of Eq.~(\ref{n30}). We start from the expansion of $\mathcal{H}$ in
powers of $\hbar$ and $\Phi$:
\beq
\label{ee1}
\mathcal{H}  =  \sum_{q = 0}^{\infty} (i \hbar)^{q} \sum_{p = 0}^{\infty}
H^{ (q) }_{ B_{p} ... B_{1} } \Phi^{ B_{1} } ... \Phi^{ B_{p} }.     
\eeq
Then the total set of structure relations for $H^{ (q) }_{ B_{p} ... B_{1} }$
is then presented as
\beq
\nonumber
&&\pa_{A} H^{ (q) }_{ B_{p} ... B_{1} }  (-1)^{ \varepsilon_{A} }  +
\sum_{ k, l, m, n, s  \geq  0 }  \delta^{q}_{ k + l + s - 1 }
\delta^{p}_{m + n} \frac{( m + s )!  ( n + s )!}{ m! s! n!}  (2)^{ - s }\times\\
\nonumber
&&\qquad\qquad\qquad\times
Sym_{ ( B_{p} ... B_{1} ) }  \left[  V^{ ( k ) }_{ A B_{p} ... B_{n +1} C_{s} ...
C_{1} }  H^{ ( l ) C_{1} ... C_{s} }_{\;\;\;\;\;\;\;\;\;\;\;\;\;\;\; B_{n} ... B_{1} }  -\right.\\
&&\qquad\qquad\qquad\quad\left.-
H^{ ( k ) }_{ B_{p} ... B_{n + 1} C_{s} ... C_{1} }
V_{A \;\;\;\;\;\;\;\;\;\;\;\;\;B_{n} ... B_{1}}^{ ( l ) C_{1}... C_{s} }
(-1)^{ \varepsilon_{A} \sum _{ i = n + 1 }^{p} \varepsilon_{ B_{ i } } }  \right]
=  0,        
\label{ee2}
\eeq
where we have the analogue  of (\ref{3}),
\beq
\label{ee3}
H^{( l ) C_{1} ... C_{s}}_{\;\;\;\;\;\;\;\;\;\;\;\;\;\;\;B_{n} ... B_{1}}  =
\eta^{ C_{1} D_{1} } ...\eta^{ C_{s} D_{s} }
H^{( l ) }_{ D_{s} ... D_{1} B_{n} ... B_{1}}
(-1)^{\sum_{ i  = 1 }^{ s} \varepsilon_{ D_{ i } }
( \sum_{ j  =  i + 1 }^{s} \varepsilon_{C_{ j }}+ 1 )}.    
\eeq

Finally, we recall the definition of the intrinsic curvilinear star product
in terms of the Weyl symbols~ \cite{FL} .   Let $\mathcal{ F}, \mathcal{G}$
be two observables,
\beq
\label{3.39}
[ T_{A},   \mathcal{F} ]_{\star}  =  0,  \quad \mathcal{F}\big|_ { \Phi = 0 }  =  F,\quad
[ T_{A},  \mathcal{G} ]  =  0,   \quad \mathcal{G}\big|_{ \Phi  = 0 }  =  G,  
\eeq
where  $T_{A}  =  P_{A}  - \mathcal{V}_{A}$  are the Abelian constraints after
conversion. The intrinsic curvilinear star product $\star_{D}$  is
defined by
\beq
\label{3.40}
F \star_{D} G  = ( \mathcal{F} \star \mathcal{G} ) \big|_{ \Phi = 0 }.  
\eeq
This star product is associative but not commutative.  The corresponding star
commutator has the form
\beq
\label{3.41}
[  F, G ] _{\star_{D}}  =  F  \star_{D}  G  -  G \star_{D}  F  (-1)^{
\varepsilon_{F} \varepsilon_{G} }.    
\eeq
In the classical limit,  the commutator (\ref{3.41})  yields exactly
curvilinear Poisson bracket (\ref{nv0.10}) ,
\beq
\label{3.42}
(  i  \hbar  )^{-1}  [  F,  G ]_{ \star_{D} } \big|_{ \hbar \rightarrow 0 }
 =  \{ F , G \}_{D}.      
\eeq
Of course, each of the symbols $\mathcal{F}$, $\mathcal{G}$  could be expanded
in power series similar to (\ref{ee1}), to satisfy the corresponding set of structure
relations (\ref{ee2}). We can then derive the corresponding expansion in $\hbar$
for intrinsic curvilinear star product (\ref{3.40}),
\beq
\nonumber
&&\qquad\qquad F \star_{D} G  =  \sum_{ q \geq 0 }  ( i  \hbar )^{q}  \sum_{ k, l, s
\geq 0 }  \delta^{q}_{ k + l + s }  s! ( 2 )^{ - s }\times\\
&&\times F^{ (k) } _{ C_{s} ... C_{1} }  \eta^{ C_{1} D_{1} }  ...  \eta^{ C_{s}
D_{s} } G^{ ( l ) }_{ D_{s} ... D_{1} }
(-1)^{  \sum_{ i  =  1 }^{s} \varepsilon_{  D_{ i } } \left( \sum_{ j  =  i  +  1
}^{s}  \varepsilon_{ C_{ j } }  +  \varepsilon_{G}  +  1  \right)  }.   
\eeq

\section{Instead of a conclusion}

We note that in this paper, we have restricted our consideration to only the
aspects closely related to the formal solution of the zero-curvature
condition for the extended symplectic potential and the corresponding
problems for the Hamiltonian and other physical observables. Here, we did
not consider aspects related to the Hilbert space of the BFF
(Batalin--Fradkin--Fradkina) converted system~\cite{BFF1,BF3,BFF3} (Fock
representation, Wick ordering, and physical states). We also did not
consider topological aspects of the global description closely related to
the structure of the bundle of symplectic spinors and the metaplectic
anomaly. A rather detailed consideration of these aspects can be found in
the excellent review by Fradkin and Linetsky~\cite{FL}. More information on the
topological aspects of geometric quantization can be found in the excellent
monograph by Karasev and Maslov~ \cite{KM}.

\section*{Acknowledgments}
\noindent
One of the authors (I.~A.~B.)~thanks Klaus Bering of Masaryk University for
the interesting discussions.
The research of I.~A.~Batalin is supported in part
by the Russian Foundation for Basic Research (Grant Nos.~14-01-00489 and
14-02-01171).  The research of P.~M.~Lavrov is supported by the
Ministry of Education and Science of the Russian Federation (Project
No.~Z.867.2014/K).

\end{document}